\documentstyle[12pt]{article}
\begin{document}
\begin{titlepage}
\vspace{1.8cm}
\begin{center}
{\bf \LARGE Combustion
of Fractal Distributions}
\\
\vspace{2cm}
{\bf Oscar Sotolongo}
\\
Department of Theoretical Physics\\
University of Havana, Havana, Cuba\\
\vspace{.3cm}
{\bf Enrique Lopez\\}
Department of Physics\\
Matanzas University, Matanzas, Cuba\\
\vspace{.5cm}

 \vspace{1cm} {\bf Abstract:} \\
 \parbox[t]{\textwidth}{
The advantages of introducing a fractal viewpoint in the field of
combustion is emphasized. It is shown that the condition for perfect
combustion of a collection of drops is the self-similarity of the
distribution  .}
 \end{center}\end{titlepage} \newpage

\section{\bf{Introduction}}

  1, The advantages of introducing geometrical viewpoints, specially fractal
  geometry in the analysis of complex phenomena is already unquestionable.
Particularly,
  such complex processes as brittle fracture have been studied experimentally
 $^{[1]}$ and theoretically$^{[2,3]}$ in order to reveal their multifractal
behaviour
  and the scaling laws present in the size distribution of the resulting
  fragments.

  Scaling is present in many fragmentation processes i.e., the
  Korcak's law for the distribution of islands, the distribution of
  icefields, and also the distribution of lunar craters, which at the same time
  reveals the distribution of meteorites (see [2] for more details). More
  recently, we have shown that scaling is also present in the breaking of
  a fluid drop$^{[4]}$ and in some regimes of atomization $^{[5]}$.

The process of fracture is important for combustion, since many combustion
chambers
burn a collection of fuel drops rather than a massive jet of fuel. Besides,
the use of water-oil emulsions in combustion devices leads to the study of
liquid
fragmentation since it has been shown$^{[6]}$ that drops of water-oil
emulsions during combustion show a disruptive behavior giving rise to
a collection of secondary droplets which increases the surface of the fluid
for the reaction with air in the process of combustion.
In all these above mentioned distributions the common characteristic is
that the cummulative number of particles with radius r (i.e, the number
of particles with radius larger than r) N(r) varies as

\begin{equation}
N(r)\sim r^{-x},
\label{eq:classic}
   \end{equation}
x being the scale exponent of the distribution. Though the geometrical model
proposed by Matsushita in [2] for fracture does not give any specific value
of x, his viewpoint permits a simple description of this process.
To obtain x one has to know
details of the dynamics of the fracture, which is a very difficult problem,
yet unsolved.
The main goal of this paper is to show the advantage of this fractal viewpoint
in the analysis of some combustion processes, specially in his application
for the analysis of the burning of water-oil emulsions.
\section{\bf{Combustion of a drop of water-oil emulsion}}
The presence of small droplets of water inside a drop of fuel
gives to the drops of water-oil emulsion a disruptive character
since they explode when burning, transforming the original drop in
a collection of small droplets that improve the process of combustion
as we pointed above.
As it was already shown$^{[4]}$ the break of a liquid drop by an
explosive process produces a collection of fragments in which their
cumulative number is given by (1). From this equation the number
of drops with radius beween r and r + dr can be found differentiating
(1), so that if we denote such distribution by n(r)dr we have
\begin{equation}
n(r)dr\sim xr^{-x-1}dr,
\label{eq:classiic}
\end{equation}
                We may suppose that the original drop, of radius A, breaks
into fragments the largest  of which  is of radius $R=\frac{A}{\beta}$ where
$\beta >1$ is a constant. Normalizing (2):

\begin{equation}
\int_{0}^{R}n(r) r^{3}dr =A^{3},
\label{eq:classico}
\end{equation}
with this we are assuming that the process of drop fracture is fast enough as
to neglect fuel consumption during fragmentation and apply mass conservation
. The normalization condition leads to the
expression for n(r):
$$n(r)=(3-x){\beta}^{3}R^{x}r^{-x-1}\eqno(4)$$

To consider the combustion of this system of fragments we may adopt
a simple model to describe the variation of the radius of an isolated drop
according to the law$^{[7]}$:
$$r_{f}^2=r_{i}^2-kt\eqno(5)$$

where $ r_{i}$ is the radius of the drop at the initial time, $r_{f}$ the
radius once elapsed the time t
and k a constant characteristic of the fuel.
We may introduce the combustion time $\tau$, so that such
drops with radius larger than $ r_{0}=\sqrt{k\tau}$ lead to a given quantity
of unburned matter, leading to soot production and waste of fuel. This
quantity can be calculated as:
     $$I=\int_{r_{0}}^{R}i(r)n(r)dr,\eqno(6) $$
     where $i(r)$ is the quantity of unburned matter given by a fragment
     of radius r, $r_{0}$ is the already introduced "critical radius"
     . The integral starts at $r_{0}$ since all drops with radius $r<r_{0}$
     will be consumed during the time $\tau$. With (5) it is easy to evaluate
     the final volume of the fragment once elapsed $\tau$. If we introduce
     the variable $\xi =\frac{r}{R}$, the quantity of unburned matter
     for one drop of radius A, expressed in units of the volume of the
     original drop $\frac{4}{3}{\pi}A^{3}$ is

$$i(A)=(3-x)\int_{\xi_{0}}^{1}\xi^{-x-1}(\xi^2-\xi_{0}^2)^{\frac{3}{2}}d\xi\eqno(7)$$
 As it can be seen, when
x approaches  3, i(A) goes to zero for any value of $ \xi_{0}$, i.e, for
any time of combustion. As x is near to 3 the combustion is improved
and for small values of $\xi_{0}$ the fuel is consumed. The case x=3
is the "ideal case" and corresponds to some kind of "ideal" or "complete"
combustion of the drop.
To interpret the case x=3 we can imagine a cube of unit length
containing a scaled distribution of drops given by (1).  Let us take
from this cube a sub-cube of length $ \lambda^{-1}, (\lambda<1)$ then
the cumulative number of drops in this sub-cube is given by
     $$N_{\lambda^{-1}}(r)\sim{\lambda^{-3}}r^{-x}, \eqno(8)$$
now we look this sub-cube with a microscope of magnification $\lambda$
such that the resolution of our observation will go from r to
$\lambda^{-1}r$, then the cumulative number observed is
$$N_{lambda{-1}}(\lambda^{-1}r)\sim\lambda^{x-3}r^{-x}.\eqno(9)$$
Comparing (9) with (1) we may conclude that x=3 means that we can
observe the same distribution of drops in any scale, i.e, in that
case the distribution is scale invariant. As the operation here
performed is essentially a renormalization transformation it can
be said that a distribution of fragments with the scale exponent
equal to the dimension of the space is the fixed point of the RG
transformation$^{[8]}$. As already was shown in [2] this fixed point
is impossible to reach.
\section{\bf{Combustion of a spray of emulsified fuel}}
The analysis of the quantity of unburned matter given by a spray
of common fuel is not different  from the preceding one for the drop. just
minor changes in notation must be made. In this respect, we will
denote as n(A) the number of atomized drops with radius A, the
total atomized volume will be denoted by by V and $\cal R$ denotes
the largest value of A. We also denote the scale exponent
for the atomization as y. Thus the quantity of unburned matter
for this case is (in units of V)
$$I_{S}=(3-y)\int_{\zeta_{0}}^{1}{\zeta}^{-x-1}[{\zeta}^2-{\zeta_{0}}^2]^{\frac{3}{2}}d{\zeta},\eqno(10)$$
where $\zeta=\frac{A}{\cal R}$ and $\zeta_{0}={\sqrt{k\tau}\over{\cal R}}.$ The
calculation of the unburned matter when
the atomized fluid is a emulsified fuel is  straightforward if we consider that
this process can be divided in two steps:  \\
-A first one in which a volume V of emulsified fuel enters the combustion
chamber with a distribution of drops characterized by a scale exponent y.\\
-A second step when each drop of initial radius A "explodes" giving a
distribution
of fragments with scale exponent x.\\
The quantity of unburned matter can be calculated in this case,
summing up all   the quantities provided by each of the drops:
$$I_{T}=\int_{\sqrt{k\tau}}^{\cal{R}}i(A)n(A)dA,\eqno(11)$$
here i(A) is given by (7). If we now express:
$$n(A)=(3-y){\gamma}{\cal{R}}^{y}A^{-y-1},\eqno(12)$$
where ${\gamma}={3V\over{4{\pi}{\cal{R}}^{3}}},$ as the distribution of
atomized drops and use (7) and (12) in (11), we obtain for
the total quantity of unburned matter in units of V:
$$I_{T}=(3-x)(3-y)\int_{\zeta_{0}}^{1}\zeta^{2-y}d\zeta
\int_{\frac{\beta\zeta_{0}}{\zeta}}^{1}\xi^{-x-1}
(\xi^{2}-\frac{\beta^{2}\zeta_{0}^{2}}{\zeta^{2}})^{\frac{3}{2}}d\zeta.\eqno(13)$$
This  expression gives us the possibility of analyzing the process of
combustion
as a function of two main factors: the scale exponents of each
distribution. It is evident that the self-similarity of the distribution
plays a major role in the improvement of combustion,leading to
the fastest consumption rate.\\
\section{\bf{Additional remarks}}
There are some curious facts emerging from this viewpoint
that we believe  important to notice:
\vskip 0.5truecm
-When evaluating the surface presented by the distribution of drops to
combustion,
we must sum the surface of each small drop and, as the drops are distributed
according to (4)
it is necessary to integrate on this distribution. If $\Sigma$ represents the
total surface of the drop distribution, then
$$\Sigma=\int_{r_0}^{R}4\pi{r}^2n(r)dr,\eqno(14)$$
the integral starts at $r_{0}$ since we are interested in the area of
the drops with size larger than $r_{0}.$ Let us denote $R=\alpha r_{0}$
with $\alpha>1$. The total surface  expressed in units of the area of the
drop is
$$\Sigma=4\pi(\frac{3-x}{2-x})(1-{\alpha}^{x-2})\eqno(15)$$
and has a finite limit when $x=2$:
$$\Sigma(x=2)\sim log\alpha,\eqno(16)$$
as can be noted, for $x=3$, $\Sigma=0$. This means that when the distribution
is completely self-similar no drops with radius larger than $r_{0}$ exist. but
as we have not fixed it, this occurs for any value of $r_{0}$ no matter
how small could it be. This corresponds with an infinite subdivision of
the drop, which leads to obvious self-similarity. The impossibility of this
value of x was obtained in [2] by another way.   \\
\vskip 0.5truecm
-For a distribution like (4) it is  always  possible to choose a value of x
such that the quantity of unburned matter given is less
than that given by a collection of small equal drops slightly larger than
$r_{0}.$ Indeed, if we represent that kind of distribution as
$n_{1}(r)\sim\delta(r_{1}-r_{0}),$
assuming that $ r_{1}=r_{0}+\epsilon$, with $\epsilon$ small,
the quantity of unburned matter produced by this kind of distribution  is
(neglecting higher order infinitesimals):
$$i_{2}=\frac{2\epsilon}{k\tau}.\eqno(17)$$
Comparing (7) and (17) it is obvious that for a fixed $\epsilon$ and $\tau$ it
is
always possible  to choose a value of x for which $i(A)<i_{2}.$ This condition
may seem shocking at first sight, but is a logical consequence of the
scaling property of the distribution.

\newpage

\section{\bf{Conclusions}}
The behavior of water-oil emulsions in the combustion process is
qualitatively different from that of conventional fuel. This is expressed in
higher consumption rates. The viewpoint here presented permits the introduction
of
the scaling exponent  as one of the important parameters to analyze combustion
processes. The fractal viewpoint seems to be natural in the analysis of drop
microexplosion and jet atomization.
\vskip 2.5truecm
{\bf Acknowledgment:} One of us (O.S.) should like to acknowledge the
hospitality
 of the International Centre for Theoretical Physics and the helpful
 discussions and comments of prof. S. Fauve.
             \newpage
             
\end {document}